\documentclass{article}

\bibstyle{abbrv}
\begin{document}
\title{What Maudlin replied to}
\author{Reinhard F. Werner}
\maketitle
\abstract{A recent post by Tim Maudlin to this archive (arXiv:1408.1828) was entitled ``Reply to Werner''. However, it was not clear to what text this was supposed to be a reply. Here I briefly provide this context, and
show that Maudlin's post is as ill-conceived as the original paper (arXiv:1408.1826).
}

\section{Background of this exchange}
In celebration of ``Fifty years of Bell's Theorem'' the Journal of Physics A carried a special issue\cite{sissue} edited by Nicolas Brunner, Otfried G\"uhne and Marcus Huber. The entire collection is now freely available under \\
{\strut \hfill\tt http://iopscience.iop.org/1751-8121/47/42\hfill\strut}\\
One contribution\cite{MaudSI}, by Tim Maudlin, was entitled ``What Bell did''. The editors found it problematic because it was very polemical, and among other things branded as a scandal that 50 years after Bell's Theorem anybody could summarize its content as excluding local realism. They therefore asked me to provide a comment explaining how Maudlin's thesis relates to the mainstream view, for which ``excluding local realism'' is not such a bad shorthand description of Bell's Theorem. I did write this comment\cite{WernerSI}, but since it was part of the editorial process of the special issue I did not immediately post it on arXiv as I normally do with all my papers. Maudlin was then invited to reply to my comment, and even before the special issue was out he jointly posted on arXiv both his paper and his reply\cite{MaudSIR} under the title ``Reply to Werner''. He hardly provided any background, and with my comment not being publicly available his submission of the reply was rather cryptic. I did get some puzzled requests, and sent the missing piece to whoever wanted it. The copyright statement I had to sign in order not to delay the publication of the special issue did not allow me to post my comment immediately, but now that the whole collection is available, it may be sufficient to just explain the sequence of events. I will also take the opportunity to straighten out some errors in Maudlin's reply.

\section{The point of the debate}

What Maudlin wants us to believe is that Bell's Theorem has locality as its only assumption, so that a violation of the inequalities directly proves that Nature itself is non-local. According to Maudlin, Bell makes no assumption of ``realism'' or (as I called it in my reply) of ``classicality'' (in short ``C''), or a hidden-variable description. This is at variance with the reading of almost everybody else. Indeed it is hard to miss the assumption in Bell's paper, as he points it out quite clearly. Maudlin's particular way of missing it is taken straight from an article\cite{scholar} in Scholarpedia, by some other authors of the Bohmian camp. In fact the only thing he adds to that paper is more polemics against anyone who does not overlook the realism assumption.

There is some tradition of trying to get rid of the hidden variable assumption in Bell's Theorem. For example, Henry Stapp has been pursuing this line in several papers. In the simplest versions he was just hiding the classicality assumption in the naive use of a logical conjunction of statements on the outcomes of incompatible observations. This is a direct petitio prinipii, and it hardly helps that the classical logic employed was modal. I recommend the reply of Abner Shimony \cite{Shimony} as a starting point to retrace this discussion from its later versions. I would agree with Shimony that Stapp's project is still a failure, and in some sense it can never succeed since algebraic quantum field theory provides an example of a theory with full relativistic signal locality and clear violations of Bell inequalities.

The approach of Maudlin and his Bohmian friends, however, is much more simplistic than Stapp's. They simply ignore the classicality assumption and deride anybody as an idiot who does not share their blind spot. Howard Wiseman \cite{WiseSI}, in the special issue in which all this was exchanged, has taken a more charitable view, explaining that the conflation of classicality and locality into ``local causality'' in a later article of Bell makes it suggestive to not see classicality as a separate assumption. But Wiseman, too, is completely aware that they {\it can} be separated. Therefore, if any kind of clarity is to be achieved in a discussion between hidden variable theorists like Maudlin and those who have learned to do Physics without classicality assumptions for the last 80 years, i.e., those who assume C and those who don't, C must be made explicit.

Whether or not assuming classicality  is a good choice is not the issue here. Therefore, at the end of the introduction of  my comment I said: ``Of course, I now have to say what this C is. I can only hope to do it well enough that Maudlin will say: 'Yes, we assume that, of course'.'' His reply shows beyond doubt that I failed. I have little hope to get through to him with a second attempt, but can at least try to explain his confusion to other readers. There are two issues, described in the following section.

\section{Comment on Maudlin's Reply}
\subsection{Maudlin's failed search for C}

The first issue is the explanation of classicality ``C''. I gave a technical definition, the simplex property, directly followed by a paragraph beginning with ``How can such a technical criterion become a far reaching postulate?''. This paragraph was apparently completely overlooked by Maudlin, who therefore misses in what sense I claim that the EPR and Bell arguments presuppose classicality. He writes, triumphantly,

{\narrower\noindent ``One can search Werner's paper high and low for this vital piece of information and it is simply nowhere to be found. Which step of that argument, exactly, does not go through if the state space of the theory is not a simplex? We are given not a shred of an indication. This is a truly remarkable circumstance. [...] about this key, central question there is literally not a word, not a breath, not a clue''. \par}

\noindent
That is quite poetically written, but in the time Maudlin took for writing that, a normal reader can easily just read on, and find and read and understand the paragraph Maudlin  was looking for ``high and low''. I was not hiding it. The point he thus missed in my explanation is that any description in terms of properties, thought to pertain to the system itself, and independent of the experimental arrangement and the choice of subsequent measurement, presupposes C. Using classical random variables, ontic states, hidden variables, and especially conditional probabilities based on those, presupposes C. And all these things are quite easy to find in the EPR and Bell arguments. Others before me apparently have tried to get the point across to Maudlin. From his main article we get the passage (also cited in my comment)

{\narrower \noindent I have heard an extremely distinguished physicist claim that Bell presupposes realism when he uses the symbol $\lambda$ in his derivation. Here is how Bell characterizes the significance of $\lambda$ ([1] 15):
``Let this more complete specification be effected by means of parameters $\lambda$. It is a matter of indifference in the following whether $\lambda$ denotes a single variable or a set, or even a set of functions, and whether the values are discrete or continuous.'' There is obviously no physical content at all in the use of the symbol $\lambda$ here. Bell makes no contentful physical supposition that can be denied. \par}

\noindent Not only the ``distinguished physicist'' is trying here, Bell himself tries to explain to his readers (and, unsuccessfully, to Maudlin) how assumption C enters. I think Bell does that fairly well. He could have said, equivalently, but in a technical language that was not his or that of his supposed readership at the time: ``Let us assume that the overall state space is a simplex, about which we will not make any further assumptions, and denote its extreme points by $\lambda$''. It is the point of this passage that a very general assumption is made, but an assumption nonetheless. Indeed it is a crucial one, and Bell seems perfectly aware of that, even if he was not entertaining the option of dropping it. And the assumption can be denied, of course. Some minimal familiarity with the literature would have shown Maudlin that it is not just crazy Werner who is doing just that. Nobody is forcing him to follow that move and, of course, it can and should be discussed. But the point of his article and the source of his scorn on the physics community (minus the few Bohmians) derives only from his own inability or unwillingness to read the basic texts of the subject.

\subsection{EPR-locality}
I cited from Maudlin's paper a passage defining ``EPR-locality''. I was aware, of course, that he uses this for a conflation of C and L, as explained by Howard Wiseman in his contribution. So I should perhaps have seen the risk of what would happen when I claimed Maudlin's definition for the operational approach. His statement, which I still find a good description of locality (L) in operational QM, was

{\narrower\noindent A physical theory is EPR-local iff according to the theory procedures carried out in one region do not immediately disturb the physical state of systems in sufficiently distant regions in any significant way.\par}

\noindent The tricky word here is ``state''. Of course, if you take it in the sense of a description by properties, as 'ontic' and pertaining just to the system by itself, you have mixed C back in, so you get that unhappy conflation. Think of a quantum state instead, a density operator for the distant system (B), as used to predict the probabilities for any further events. From such further events you would anyhow have to tell that a disturbance had happened. And indeed in the operational approach no prediction about B changes when or if a measurement or other procedure is carried out on A. This independence is built into the structure of quantum theory. This is also the same as the no-signalling condition and the possibility of tracing out system A, getting a reduced state for B, which does not change (and so is undisturbed) whatever happens just to A.

I tried to make clear that I was not talking about the hidden variable sort of state by bringing in the ontic-epistemic distinction. That was perhaps another didactic mistake because this is often done inside the hidden-variable camp, and possibly got Maudlin on a wrong track. What apparently confused him further is the obvious fact that if you condition on the outcome of a measurement on A, you get a modified state for B\footnote{By the way, conditional states can be used to express and analyze the EPR argument, with C brought in explicitly as the assumption that it is possible to interpret the conditioning in terms of filtering from a distribution of antecedent ``hidden ontic states''. This is expressed in the notion of ``steering'' and is further explained in my main article in the special issue.}. That is just another way to look at correlation, but never, not even in classical probability, can this be confused with a physical disturbance. The state change only becomes effective when the results from the two labs are brought together and are jointly analyzed, which can happen centuries later. Bohmians like Maudlin tend to confuse such changes in distributions with a change in the world, because the notions of states and wave functions are reified, and considered as some real thing out there. If a state change by conditioning on a remote observation were to count as ``disturbing the physical state of the system in a significant way'' then any classical probabilistic theory allowing correlations would be not EPR-local. Attributing such a silly notion to EPR would completely miss the new points raised in that paper, and could only be seen as a deliberate insult to Einstein, Podolsky, and Rosen. I can only hope this is not intended by Maudlin, but then I cannot see how QM fails his criterion.

\section{Summary}
As Maudlin rightly points out, nobody in this debate from the times of EPR onwards (except some parapsychologists\cite{Schmidt}) ever claimed signalling on mere correlations would work. Nevertheless,  different consequences have been drawn from this consensus: Operational quantum mechanics has built it into its very structure (i.e., assuming L, hence dropping C). Bohmian theory has {\it chosen} to stick to C which then forces the negation of L at the level of its hidden variables. This and only this is the meaning of the term ``non-locality'' in quantum mechanics.

Of course, none of this is new. Like many of my colleagues, I expect my students to know it, and indeed many interested lay people do. No doubt there are also many misunderstandings around, and clearing those up is a very laudable undertaking. Unfortunately, Maudlin's articles are  backward steps towards this goal.


\end{document}